\documentclass[pre,twocolumn,preprintnumbers,amsmath,amssymb,nopacs,nofootinbib,floatfix]{revtex4}

\usepackage{graphicx,bm,xcolor, bbold, physics, enumitem}
\usepackage[normalem]{ulem}
\usepackage{hyperref}

\makeatletter
\def\graphicscale{\twocolumn@sw{0.3}{0.4}}
\def\graphicthreescale{\twocolumn@sw{0.3}{0.4}}

\begin{document}

\title{Color-flavor reflection in the continuum limit of \\ two-dimensional lattice gauge
    theories with scalar fields}

\author{Claudio Bonati}
\affiliation{Dipartimento di Fisica dell'Universit\`a di Pisa
        and INFN, Largo Pontecorvo 3, I-56127 Pisa, Italy}

\author{Alessio Franchi}
\affiliation{Dipartimento di Fisica dell'Universit\`a di Pisa
        and INFN, Largo Pontecorvo 3, I-56127 Pisa, Italy}

\date{\today}

\begin{abstract}
We address the interplay between local and global symmetries in determining the
continuum limit of two-dimensional lattice scalar theories characterized by
$SO(N_c)$ gauge symmetry and non-Abelian $O(N_f)$ global invariance. We argue
that, when a quartic interaction is present, the continuum limit of these model 
corresponds in some cases to the gauged non-linear $\sigma$ model field theory associated with
the real Grassmannian manifold $SO(N_f)/(SO(N_c)\times SO(N_f-N_c)$), which
is characterized by the invariance under the color-flavor reflection
$N_c\leftrightarrow N_f-N_c$.  Monte Carlo simulations and Finite-Size Scaling
analyses, performed for $N_f=7$ and several values of $N_c$, confirm the
emergence of the color-flavor reflection symmetry in the scaling limit, and
support the identification of the continuum limit.
\end{abstract}

\maketitle


\section{Introduction}

Global and local symmetries are cornerstones of our understanding of many
different physical phenomena, ranging from  high energy physics to many-body
systems in condensed matter physics \cite{W_the_q, A_basic}. Global symmetries
characterize fundamental features of a physical model, like its phase diagram
and energy spectrum \cite{Z_quant}. Gauge symmetries, instead, constraint
the form of the possible interactions, are responsible for the presence of
Higgs phases, and are necessary to describe some emergent phenomena in
condensed-matter systems \cite{W_quantum, S_topol}. 

The nature of the interplay between local and global symmetries at a phase
transition is a long-standing issue \cite{EB_broken, H_broke, GHK_global,
PW_remarks, ARW_color}, which however is still not completely clarified.
Two-dimensional scalar models with local and global symmetries are perfect
playgrounds to investigate this topic, since their critical properties can be
studied by means of numerical simulations requiring only a moderate
computational effort. In particular, recent works have considered
two-dimensional scalar models with orthogonal local symmetries
\cite{BFPV_asymp, BFPV_berez}, and models with unitary local symmetries and
scalar matter transforming in different representations of the gauge group
\cite{BPV_unive, BFPV_two-d}. 

According to the Mermin-Wagner theorem \cite{MW_absen}, in two
dimensional systems the critical behavior associated with the breaking of a
continuous global symmetry can be observed only in the zero temperature limit.
This obviously remains true also for gauge models, but in this case two
different classes of low-temperature behavior can a priori be expected to
emerge, depending whether the gauge degrees of freedom are or are not critical 
at the transition.

For most of the two-dimensional gauge models studied so far (see e.g.
\cite{BFPV_asymp, BFPV_berez, BPV_unive}) gauge degrees of freedom do not
develop long range correlations at the transition. In these cases gauge
symmetry does not seem to play a pivotal role in determining the
low-temperature universal critical behavior: the size of the gauge group (i.e.
$N_c$) does not influence the critical low-temperature regime, just like the
specific matter field representation. Also in this case, however, gauge
symmetry plays in fact a fundamental role: that of selecting the critical
degrees of freedom, preventing non-gauge invariant observables from developing
critical correlations. Gauge fields are instead expected to become critical
for some values of the parameters of the model studied in \cite{BFPV_two-d},
where this fact was attributed to the presence of a quartic scalar field
interactions (see \cite{SSST-19,SPSS-20, BFPV_phase} for the analogous three-dimensional
case).

The aim of this paper is to extend the results of \cite{BFPV_asymp, BFPV_two-d}, by studying
a lattice model characterized by $SO(N_c)$ local invariance, multicomponent
scalar fields transforming in the fundamental representation of the color
group, and non-Abelian $O(N_f)$ global symmetry together with a quartic
interaction. This field content is indeed better suited than the one used in
\cite{BFPV_two-d} to investigate a peculiar symmetry between $N_c$ and $N_f$,
that we expect to emerge in the continuum limit. The origin of this symmetry is
the following: we will argue the continuum limit of the model studied in this
paper to be (at least in some regions of the parameter space) the Non-Linear
$\sigma$ Model (NL$\sigma$M) field theory associated with the Grassmannian
manifold $SO(N_f)/(SO(N_c)\times SO(N_f-N_c)$). Since this manifold is clearly
invariant under the exchange $N_c\leftrightarrow N_f - N_c$, we expect this
``color-flavor reflection symmetry'' to characterize also the low-temperature
critical behavior of the gauge model.  A numerical verification of the
emergence of this symmetry will strongly support the identification of the
continuum limit with the Grassmann NL$\sigma$M field theory.

The paper is organized as follows: in Sec.~\ref{sec_lattice_model} we present
the lattice model. In Sec.~\ref{sec_minima} we discuss its minimum-energy
configurations controlled by the quartic interaction, and identify the
different continuum limits exhibited by the lattice model in the space of
parameters. In Sec.~\ref{sec_obs_and_fss} we introduce the observables and the
Finite-Size Scaling (FSS) theory used in actual numerical analyses. In
Sec.~\ref{sec_numerical_results} we  present our numerical results, in
Sec.\ref{sec_conclusions} we finally draw our conclusions and present possible
outlooks of our work. Some technical details concerning the algorithms and
the Monte Carlo simulations are reported in App.~\ref{appendix}.

\section{The lattice model}
\label{sec_lattice_model}

The matter field of the lattice model studied in this work is a multicomponent
real scalar field $\Phi^{i\alpha}_{\bm x}$, where $i=1,..,N_c$ and
$\alpha=1,..,N_f$ stand for the ``color'' and the ``flavor'' indices respectively, and
${\bm x}=(x_1, x_2)$ denotes the position on the lattice. For the sake of the
simplicity we will adopt the unit-length (London limit) constraint $\Tr
\Phi^t_{\bm x}\Phi_{\bm x}=1$, which is not expected to alter the critical
behavior of the model.

To define the lattice Hamiltonian, we start from the most general quartic
interaction for the field $\Phi^{i\alpha}$ compatible with the $O(N_c)\times
O(N_f)$ global invariance, i.e. from the lattice NL$\sigma$M with Hamiltonian
\begin{equation}
  H_\sigma = - N_f\sum_{{\bm x}, \mu>0}\Tr\Phi^t_{{\bm x}}\Phi_{{\bm x}+\hat{\mu}} + 
  w \sum_{{\bm x}}\Tr\Phi_{\bm x}^t\Phi_{\bm x}\Phi_{\bm x}^t\Phi_{\bm x}\,,
  \label{ham_sigma}
\end{equation}
where $\hat{\mu}=\hat{1}, \hat{2}$ denotes a lattice vector, the
coefficient of the first term has been normalized to $N_f$, and periodic boundary
conditions in all directions are assumed. The lattice gauge model is obtained
by gauging the color degrees of freedom of the matter field, coupling
them to a link variable $U_{{\bm x},\mu}\in SO(N_c)$ following Wilson's
prescription \cite{W_confi}. 

The complete Hamiltonian can thus be written in the form
\begin{equation}
    H = H_K(\Phi_{\bm x}, U_{{\bm x}, \mu}) + V(\Phi_{\bm x}) + H_G(U_{{\bm x},\mu})\,.
    \label{ham}
\end{equation}
In this expression $H_K(\Phi, U)$ and $V(\Phi)$ represent the scalar field
kinetic energy and the quartic interaction respectively, which are defined as
follows
\begin{align}
  H_K(\Phi_{\bm x}, U_{{\bm x},\mu}) &= - N_f\sum_{{\bm x}, \mu>0} 
  \Tr \Phi^t_{{\bm x}} U_{{\bm x},\mu}\Phi_{{\bm x}+\hat{\mu}} \ ,
  \label{H_K}\\
  V(\Phi_{\bm x}) &= w \sum_{{\bm x}} \Tr \Phi^t_{\bm x} \Phi_{\bm x} \Phi^t_{\bm x} \Phi_{\bm x}\,.
  \label{V}
\end{align}
The last term $H_G(U_{\bm x, \mu})$ in Eq.~\eqref{ham} represents instead the
kinetic term of the gauge field, and it is written by means of the plaquette
operator $\Pi_x$ in the usual Wilson form:
\begin{align}
    H_G(U_{\bm x}) &= - \frac{\gamma}{N_c}\sum_{\bm x} \Tr \Pi_{\bm x}\ ,\\
    \Pi_{\bm x} &= U_{{\bm x},1}U_{{\bm x}+\hat{1},2}U^t_{{\bm x}+\hat{2},1}U^t_{{\bm x},2}\,.
\end{align}
The model is characterized by a gauge $SO(N_c)$ invariance under the local transformation
\begin{equation}
    \Phi_{\bm x} \mapsto W_{\bm x}\Phi_{\bm x}, \quad U_{{\bm x},\mu}\mapsto W_{\bm x} 
    U_{{\bm x},\mu} W^t_{{\bm x}+\hat{\mu}}\,,
\end{equation}
where $W_{\bm x} \in SO(N_c)$, and by a global $O(N_f)$ invariance under $\Phi_{\bm
x}\mapsto\Phi_{\bm x}M$, with $M\in O(N_f)$. 
The partition function of the statistical model is finally defined by 
\begin{equation}
    Z = \sum_{\{\Phi, U\}} e^{-\beta H}\ ,
\end{equation}
where $\beta$ plays the role of the inverse temperature.

The Hamiltonian in Eq.~\eqref{ham} has several limiting cases which correspond
to known and already studied models. For $w=0$ and $\gamma=\infty$ it
trivially reduces to the standard $O(N_cN_f)$ model, while for $w=0$ and
finite $\gamma$ it reduces to the model studied in \cite{BFPV_asymp}. The case
$N_c=2$ for $w=0$ is quite peculiar, since the global symmetry is not
$O(N_f)$ but $U(N_f)$, see \cite{BPV_three}. Finally, as will be clear from
the analysis of the minimum-energy configurations carried out in the next section,
when $N_f>N_c$ the Hamiltonian reduces in the limit $w=\infty$ and $\gamma=\infty$ to that of the so called Stiefel models, which are the NL$\sigma$Ms defined on the
homogeneous spaces $SO(N_f)/SO(N_f-N_c)$ \cite{Z_phase, KZ_stief, L_phase}. 

In this work we will mainly focus on the case $\gamma=0$, as we do not expect
the gauge coupling to play a relevant role at criticality in two dimensions
(apart from crossover effects in the limit $\gamma\to\infty$), just as in all
previously studied cases \cite{BFPV_asymp, BFPV_berez, BPV_unive, BFPV_two-d}.

\section{Minimum-energy configurations and continuum limits}
\label{sec_minima}

In this section we identify the minimum-energy configurations of the gauge
model in Eq.~\eqref{ham}, which are expected to be the ones characterizing the
low-temperature ($\beta\to+\infty$) critical behavior of the model. The
minimum-energy configurations are selected by the quartic interaction, and
their determination uses arguments that partially retrace those used in
Refs.\cite{BFPV_phase, BFPV_two-d}.  

By means of the singular value decomposition, the scalar field matrix
$\Phi_{\bm x}^{i\alpha}$ can be rewritten as follows
\begin{equation}
    \Phi^{i\alpha}_{\bm x} = \sum_{j=1}^{N_c}\sum_{\delta=1}^{N_f}C^{ij}_{\bm x}
    D^{j\delta}_{\bm x}F^{\alpha\delta}_{\bm x}\,,
    \label{svd}
\end{equation}
where $C$ and $F$ are two square matrices in $O(N_c)$ and $O(N_f)$
respectively, $D=\mathrm{diag}\{s_1,..,s_q\}$ is a $N_c\times N_f$ matrix with
non-negative diagonal entries and $q\equiv\text{min}[N_c, N_f]$. 
The fixed-length constraint becomes 
\begin{equation}
    \Tr \Phi^t_{\bm x}\Phi_{\bm x} = \displaystyle \sum_{i=1}^{q}s_i^2=1\ ,
\end{equation}
and the the quartic potential term is  equal to
\begin{equation}
    \Tr \Phi^t_{\bm x}\Phi_{\bm x}\Phi^t_{\bm x}\Phi_{\bm x} = 
    \displaystyle \sum_{i=1}^{q}s_i^4 \ .
\end{equation}

The explicit form of the minimum-energy configurations is
fixed by the sign of the quartic coupling $w$:
\begin{enumerate}[label=(\Roman*)]
    \item $s_1=1$\ , $s_2=s_3=\ldots=s_q=0$  \hspace{0.5cm} if $w<0$
    \label{minima1}
    \item $s_1=\ldots=s_q=1/\sqrt{q}$ \hspace{1.98cm}    if $w>0\,. $
    \label{minima2}
\end{enumerate}
The residual symmetry of these minima determines the low-temperature critical
behavior of the model, which is thus expected to be different for positive
or negative values of the coupling $w$. The case $w=0$ requires
special attention and, apart from a few cases in which analytical results can
be obtained (see \cite{BFPV_phase, BFPV_two-d}), we have to rely on numerical
simulations for the identification of the critical behavior when $w=0$. 

When configurations of type \ref{minima1} dominate the partition function (i.e.
for $w<0$), the continuum limit is expected to be the same as the
$\mathbb{R}P^{N_f-1}$ model \cite{BFPV_two-d, BFPV_asymp}. The simplest way to
understand this fact is to note that, by using configurations of type
\ref{minima1}, it is possible to construct the gauge invariant composite field
$B_{\bm x}=\Phi^t_{\bm x}\Phi_{\bm x}$, which behaves as a rank-1 real space
projector (i.e. $B_{\bm x}^2=B_{\bm x}=B_{\bm x}^t$ and $\Tr B_{\bm x}=1$).
Indeed the $\mathbb{R}P^{N_f-1}$ model describes the dynamics of a rank-$1$
projector $P_{\bm x}$ in an $N_f$ dimensional space.  By introducing the
unit-length vector ${\bm S}_{\bm x}$ using the definition
$P^{\alpha\delta}_{{\bm x}}= S^{\alpha}_{\bm x}S^{\delta}_{\bm x}$, the action
of the $\mathbb{R}P^{N_f-1}$ model can be written as
\cite{BFPV_asymp_rpn, H_onand, NWS_quest, CHHR_natur} 
\begin{equation}
    H_{RP} = - \sum_{{\bm x}, \mu} \Tr P_{\bm x}P_{{\bm x}+\hat{\mu}}=
    -\sum_{{\bm x}, \mu} \big( S_{\bm x} \cdot S_{{\bm x} + \hat{\mu}} \big)^2\,,
\end{equation}
which displays the two main features of the $\mathbb{R}P^{N_f-1}$ universality
class: $O(N_f)$ global invariance and $\mathbb{Z}_2$ local invariance. We thus
expect the $SO(N_c)$ symmetry not to play any significant role in
the continuum limit of the model for $w<0$. Moreover in \cite{BFPV_asymp} it
was numerically shown that the same is also true for the case $w=0$, and for
all the values of $\gamma$ investigated.

We can now discuss the case in which the dominant configurations are those of
type \ref{minima2}. Using an argument analogous to that used for the unitary
groups in \cite{BFPV_phase}, it is possible to show that if $N_f\leq N_c$ there
is no residual global symmetry for the field $\Phi_{\bm x}$ in this case, hence
no critical behavior at all is expected in the low-temperature limit when $w>0$
(see \cite{BFPV_two-d} for a numerical check). For this reason, in the
following, we will assume $N_f>N_c$, hence $q=N_c$. As for the case of type
\ref{minima1} minima, it is convenient to introduce a gauge invariant composite
field, which is now $\widetilde{B}_{\bm x}=N_c\Phi^t_{\bm x}\Phi_{\bm x}$. It
is indeed simple to show, using the explicit form of the type \ref{minima2}
minima, that $\widetilde{B}_{\bm x}^2=\widetilde{B}_{\bm x}=\widetilde{B}_{\bm
x}^t$ and $\Tr \widetilde{B}_{\bm x}=N_c$, hence $\widetilde{B}_{\bm x}$ is a
rank-$N_c$ real space projector.  The lattice NL$\sigma$M written by using
$\widetilde{B}_{\bm x}$, whose Hamiltonian is
\begin{equation}
H_G=- \sum_{{\bm x}, \mu} \Tr \widetilde{B}_{\bm x}\widetilde{B}_{{\bm x}+\hat{\mu}}\ ,
\end{equation}
is a possible discretization of the NL$\sigma$M field theory associated with
the real Grassmannian manifolds
\begin{equation}\label{G_symm}
SO(N_f)/(SO(N_c)\times SO(N_f-N_c))\ ,
\end{equation}
where we have neglected discrete subgroups that, as usual, are not expected to play
any role in the zero temperature limit.

The same conclusion about the global symmetry breaking pattern for $w>0$ can be
reached also in another way, by using as effective model the Hamiltonian
$H_K(\Phi_{\bm x}, U_{{\bm x},\mu})$, with the field $\Phi_{\bm x}$ restricted
to be of the form \ref{minima2}. Introducing the rescaled field
$\widetilde{\Phi}_{\bm x} = \sqrt{N_c}\Phi_{\bm x}$, the effective model
Hamiltonian is written as
\begin{equation} \label{HKwinf}
    H_K^{w\to\infty} = - \frac{N_f}{N_c}\sum_{{\bm x}, \mu} \Tr \widetilde{\Phi}_{\bm x}^t 
    U_{{\bm x}, \mu} \widetilde{\Phi}_{{\bm x}+\mu}\,.
\end{equation}
By using essentially the same arguments discussed in \cite{BFPV_phase} for the
unitary group case, one obtains again for the global invariance group of
$H_K^{w\to\infty}$ the expression in Eq.~\eqref{G_symm} (again up to discrete groups).

Two-dimensional Grassmannian NL$\sigma$M field theories have been introduced in
Refs.~\cite{P_nonlinear, BHZ_gener}, are known to be asymptotically free
\cite{BHZ_gener, Z_quant} and their $\beta$-functions are known up to 4-loops
in dimensional regularization \cite{W_fourl}. Since Grassmann manifolds are
invariant under the color-flavor reflection $N_c\leftrightarrow N_f-N_c$, our
study of the minimum-energy configurations leads us to expect this symmetry to
emerge in the critical low-temperature behavior of the lattice model in
Eq.~\eqref{ham} when $w>0$. We explicitly note that this
symmetry is present also when $w<0$, however in that case it is realized in a
somehow trivial way: the critical behavior is expected to always be that of the
$\mathbb{R}P^{N_f-1}$ model, for all $N_c$ values. In
Sec.~\ref{sec_numerical_results} we will provide numerical evidence that, for
$w>0$, the universal FSS curves of the lattice model considered in this paper
do depend on $N_c$, and that the color-flavor reflection symmetry is realized. 

We close this discussion by noting that the second approach used above to
identify the invariance group when $w>0$, and in particular the effective
Hamiltonian in Eq.~\eqref{HKwinf}, is especially convenient to clarify the
relation of the model we are studying with the Stiefel models. In the limit
$\gamma\to\infty$ we have $U_{{\bm x},\mu}\to 1$ up to gauge transformations
(at least in the thermodynamic limit). Moreover it is immediate to verify that
for type \ref{minima2} minima the following relation holds true
\begin{equation}
   \widetilde{\Phi}_{\bm x}\widetilde{\Phi}_{\bm x}^t = \mathbb{1}_{N_c\times N_c}\ , 
\end{equation}
where $1_{N_c\times N_c}$ denotes the $N_c\times N_c$ identity matrix (note
that this relation is gauge invariant although color indices are not
contracted). For $\gamma\to\infty$ the Hamiltonian $H_K^{w\to\infty}$ in
Eq.~\eqref{HKwinf} thus reduces, up to an irrelevant multiplicative factor that
can be reabsorbed in the normalization of $\beta$, to that of the Stiefel model
$V_{(N_f,N_c)}$, which is usually written as \cite{Z_phase, KZ_stief, L_phase}
\begin{equation}
  H_S = - N_f \sum_{{\bm x}, \mu} \Tr \pi^t_{\bm x} \pi_{{\bm x} + \mu}\,,
  \label{ham_stiefel}
\end{equation}
where $\pi_{\bm x}$ are $N_c\times N_f$ real matrices (with $N_f>N_c$) satisfying the constraints
\begin{equation}
  \pi_{\bm x}\pi_{\bm x}^t=\mathbb{1}_{N_c\times N_c}\ .
\end{equation} 
The continuum limit of this model is described by the NL$\sigma$M field theory having the manifold $SO(N_f)/SO(N_f-N_c)$ as target space, and which is associated
to the symmetry breaking pattern $SO(N_f)\to SO(N_f-N_c)$ (see e.g.
\cite{Z_phase} for more details).

\begin{figure}
    \centering
    \includegraphics[width=0.95\columnwidth, clip]{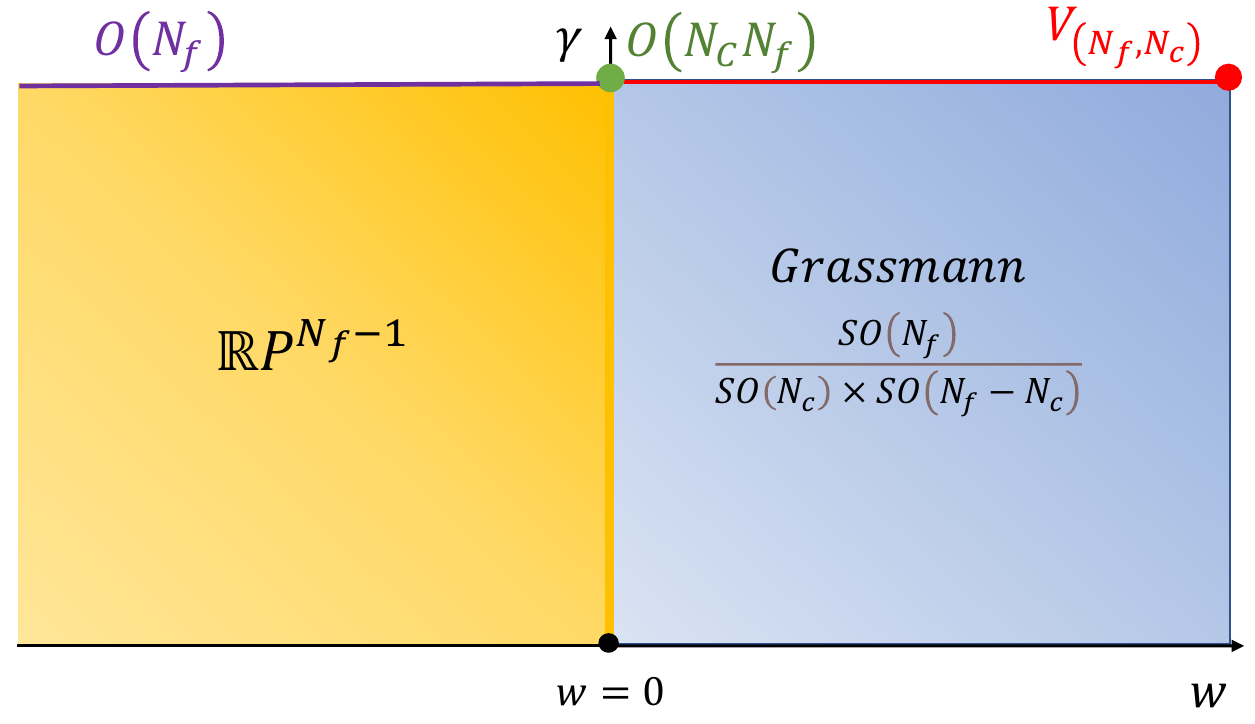}
    \caption{Graphical representation of the different low-temperature
    continuum limits expected from the study of the minimum-energy configurations when $N_f>N_c$.
    Minima of type \ref{minima1} ($w<0$) and \ref{minima2} ($w>0)$ are associated
    to the $\mathbb{R}P^{N_f-1}$ and Grassmannian models respectively.  The case 
    $w=0$ is numerically found to have the same asymptotic behavior of the $w<0$ cases
    (see \cite{BFPV_asymp}).
    For $\gamma\to+\infty$ we recover the $O(N_f)$ vector model when $w<0$, the
    $O(N_cN_f)$ vector model when $w=0$ and the Stiefel model $V_{(N_f, N_c)}$ when
    $w>0$. It $N_f\le N_c$ no critical behavior is expected for $w>0$.}
    
    \label{fig_phase_diagram}
\end{figure}

A summary graph of the different low-temperature behaviors expected on the basis of
the residual symmetry of the minimum-energy configurations is shown in
Fig.~\ref{fig_phase_diagram}.

\section{Observables and Finite-Size Scaling}\label{sec_obs_and_fss}

To characterize the critical behavior of the lattice models we use Monte Carlo
simulations and Finite-Size Scaling (FSS) techniques \cite{FB_scali,
P_finit, PV_criti}. We focus on the bilinear local operator $Q_{\bm x}$
\begin{equation}
  B^{\alpha\beta}_{{\bm x}}=\sum_{i=1}^{N_c}\Phi^{i\alpha}_{\bm x}\Phi^{i\beta}_{\bm x},\quad 
  Q^{\alpha\beta}_{\bm x} = B^{\alpha\beta}_{{\bm x}} - \frac{\delta^{\alpha\beta}}{N_f}\,,
\end{equation}
and, more specifically, on some Renormalization Group (RG) invariant quantities
associated to this operator. Starting from the two point correlation function 
\begin{equation}
    G({\bm x}-{\bm y}) = \expval{\Tr Q_{\bm x} Q_{\bm y}}\,,
\end{equation}
we can define the second-moment correlation length
\begin{equation}
    \xi^2 = \frac{1}{4 \sin^2(\pi/L)}\frac{\widetilde{G}({\bm 0}) - 
            \widetilde{G}({\bm p}_m)}{\widetilde{G}({\bm p}_m)}\,,
\end{equation}
where $\widetilde{G}({\bm p}) = \sum_{\bm x}e^{i{\bm p}{\bm x}}G({\bm x})$ is
the Fourier transform of $G({\bm x})$ and ${\bm p}_m=(2\pi/L, 0)$ is the
minimum momentum on the lattice. The first RG invariant quantity we consider is
the ratio between the second-moment correlation length and the lattice
size 
\begin{equation}
  R_\xi \equiv \xi/L\ ,
\end{equation}
and lattices with equal extent along the two directions will always be adopted. The second 
RG invariant quantity we use is the quartic Binder cumulant $U$, defined by
\begin{equation}
  U = \frac{\expval{\mu_2^2}}{\expval{\mu_2}^2}\,,\quad 
  \mu_2 = \frac{1}{L^4}\sum_{{\bm x},{\bm y}}\Tr Q_{\bm x}Q_{\bm y}\ .
\end{equation}

Since $R_{\xi}$ is found to be a monotonic function of the temperature, it is
convenient to use $R_{\xi}$ instead of $\beta$ to parametrize the Binder
cumulant $U$, writing $U(\beta, L)=U(R_\xi, L)$. Indeed, since both $R_{\xi}$ and $U$ are
RG-invariant quantities, the curve $U(R_{\xi}, L)$ is expected to approach in the
FSS limit (i.e. for $L\to\infty$ at fixed $R_{\xi}$) a universal scaling
curve 
\begin{equation}\label{FSS_relation}
U(R_{\xi},L)\stackrel{\mathrm{FSS}}{\longrightarrow} \mathcal{U}(R_\xi)\ ,
\end{equation}
where $\mathcal{U}(R_{\xi})$ depends on the universality class of the
transition, the lattice boundary conditions and aspect ratio. Scaling
corrections of the form $L^{-2}\log^p L$ are expected, as for all
asymptotically free theories (see e.g. \cite{CP_corre} for a detailed analysis of the $O(N)$
models).

In the next section we will use Eq.(\ref{FSS_relation}) to investigate whether
two different lattice systems share the same universality class. We will also
present some numerical results obtained for the Stiefel models in
Eq.~\eqref{ham_stiefel}. The observables used for the Stiefel models can be
obtained from the ones defined in this section by the replacement $\Phi_{\bm
x}\to \pi_{\bm x}/\sqrt{N_c}$.

\section{Numerical results}\label{sec_numerical_results}

In this section we discuss the numerical results obtained by simulating the
model defined in Eq.(\ref{ham}). We carried out Monte Carlo simulations
and FSS analyses for fixed $N_f=7$ and several values of $N_c$, to
check the emergence of the color-flavor reflection symmetry $N_c\leftrightarrow
N_f-N_c$ in the asymptotic low-temperature critical behavior when $w>0$. Technical details
concerning Monte Carlo simulations are postponed to the App.~\ref{appendix}.

\begin{figure}[tb]
    \centering
    \includegraphics[width=0.95\columnwidth, clip]{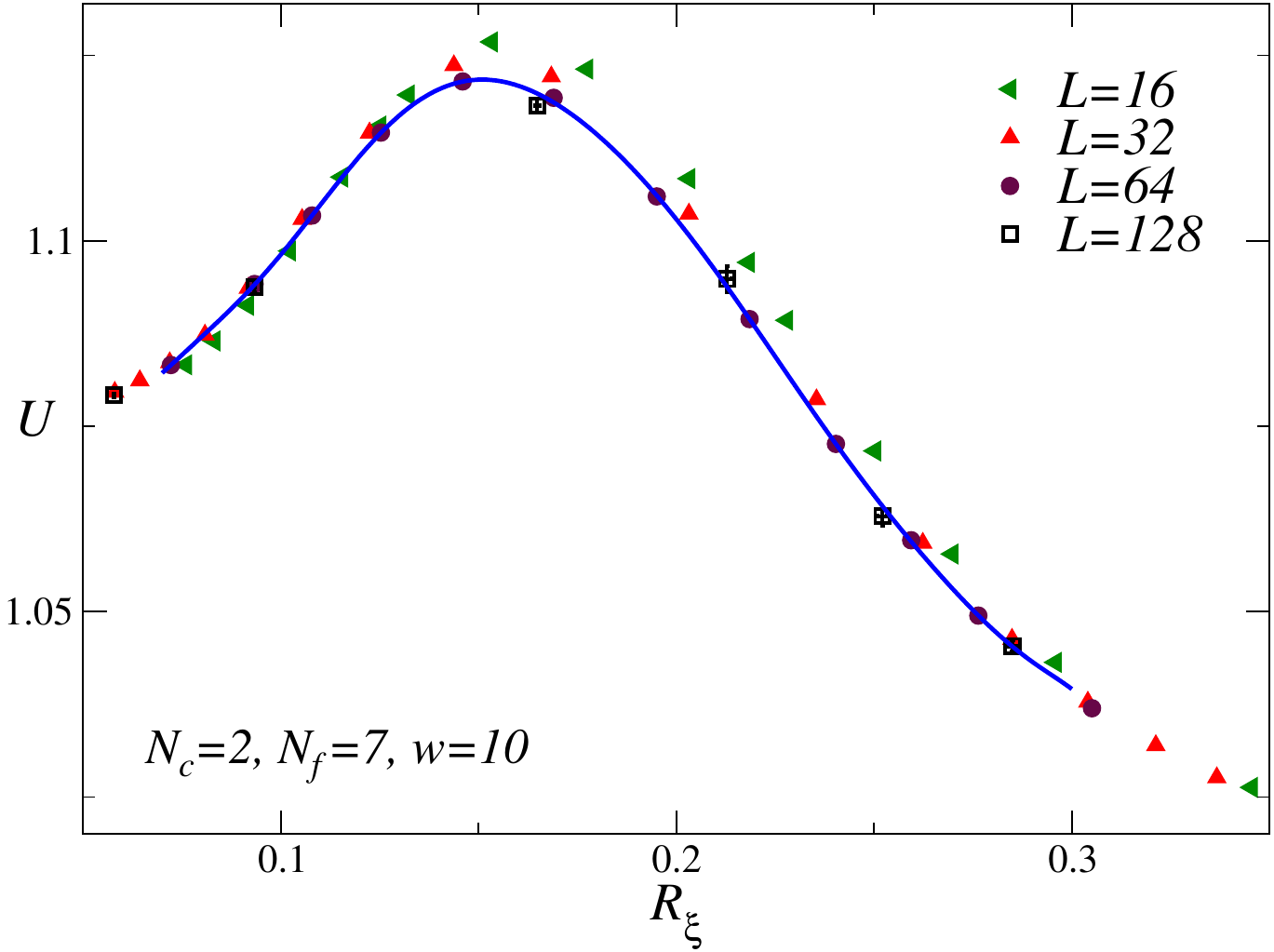}
    \includegraphics[width=0.95\columnwidth, clip]{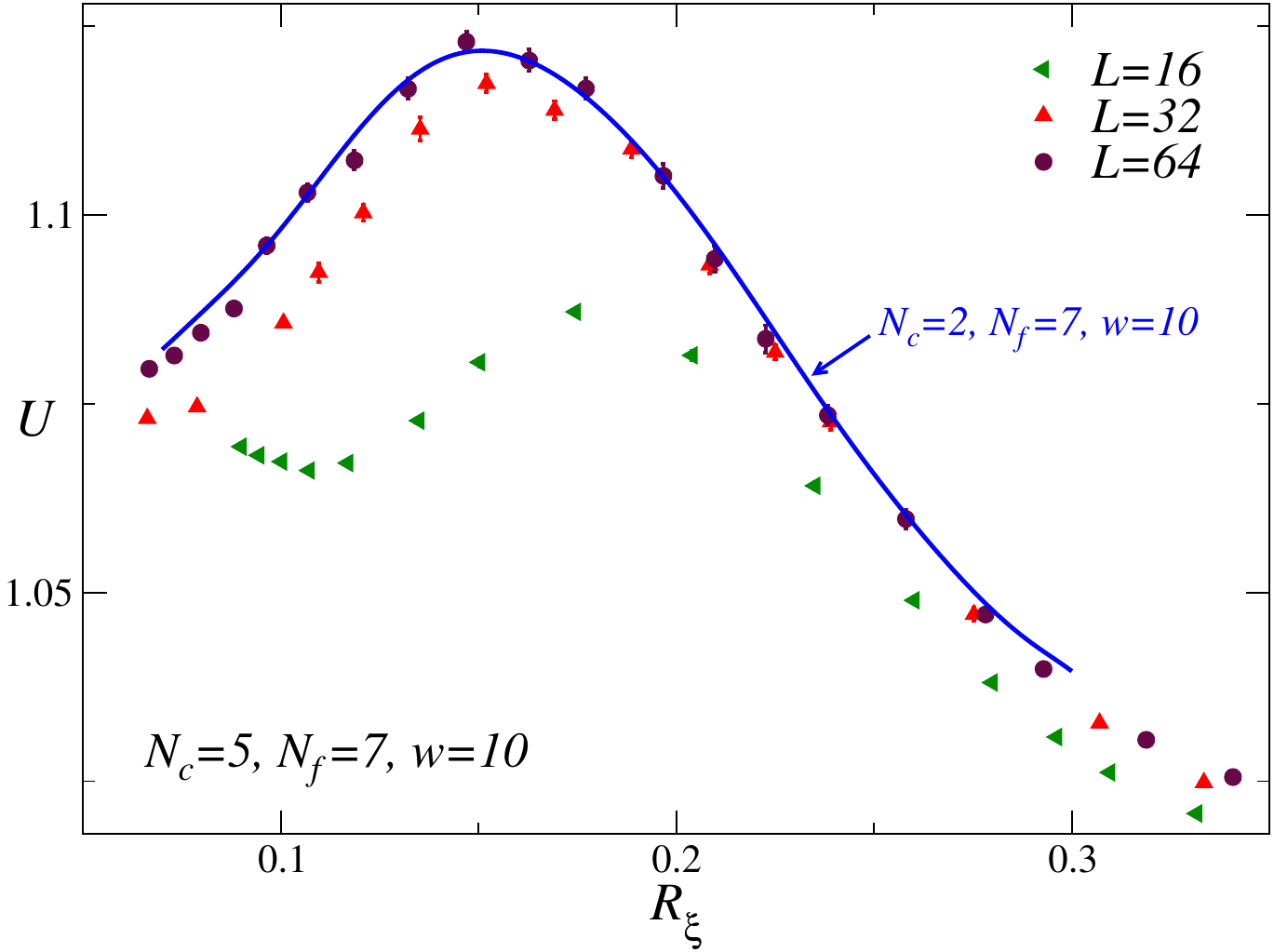}
    \caption{Top: Binder cumulant $U$ versus $R_\xi$ for the case $N_c=2,
N_f=7$ and quartic coupling $w=10$ ($\gamma=0$). The solid line is a polynomial interpolation
of the $L=64$ and $L=128$ data, and it is our estimate of the asymptotic curve
$\mathcal{U}(R_\xi)$ for the Grassmannian $SO(7)/(SO(2)\times SO(5))$ 
(note that $L=64$ and $L=128$ data are consistent with each other). Bottom:
Binder cumulant $U$ versus $R_\xi$ for the case $N_c=5,
N_f=7$ and quartic coupling $w=10$ ($\gamma=0$). The solid line is the estimate of 
$\mathcal{U}(R_\xi)$ obtained from $N_c=2$ data.}
    \label{fig2c7fw10}
\end{figure}

\begin{figure}[tb]
    \centering
    \includegraphics[width=0.95\columnwidth, clip]{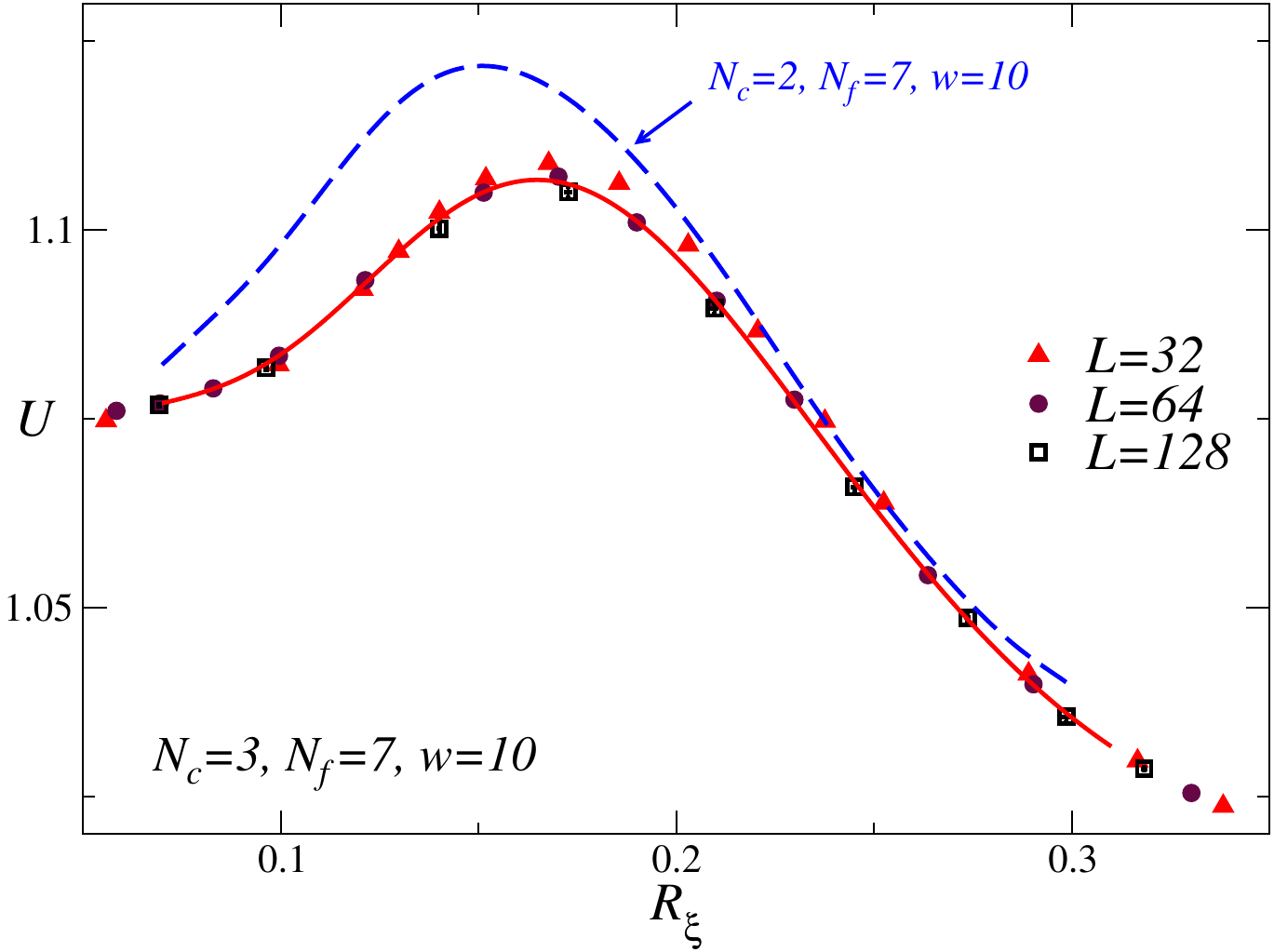}
    \includegraphics[width=0.95\columnwidth, clip]{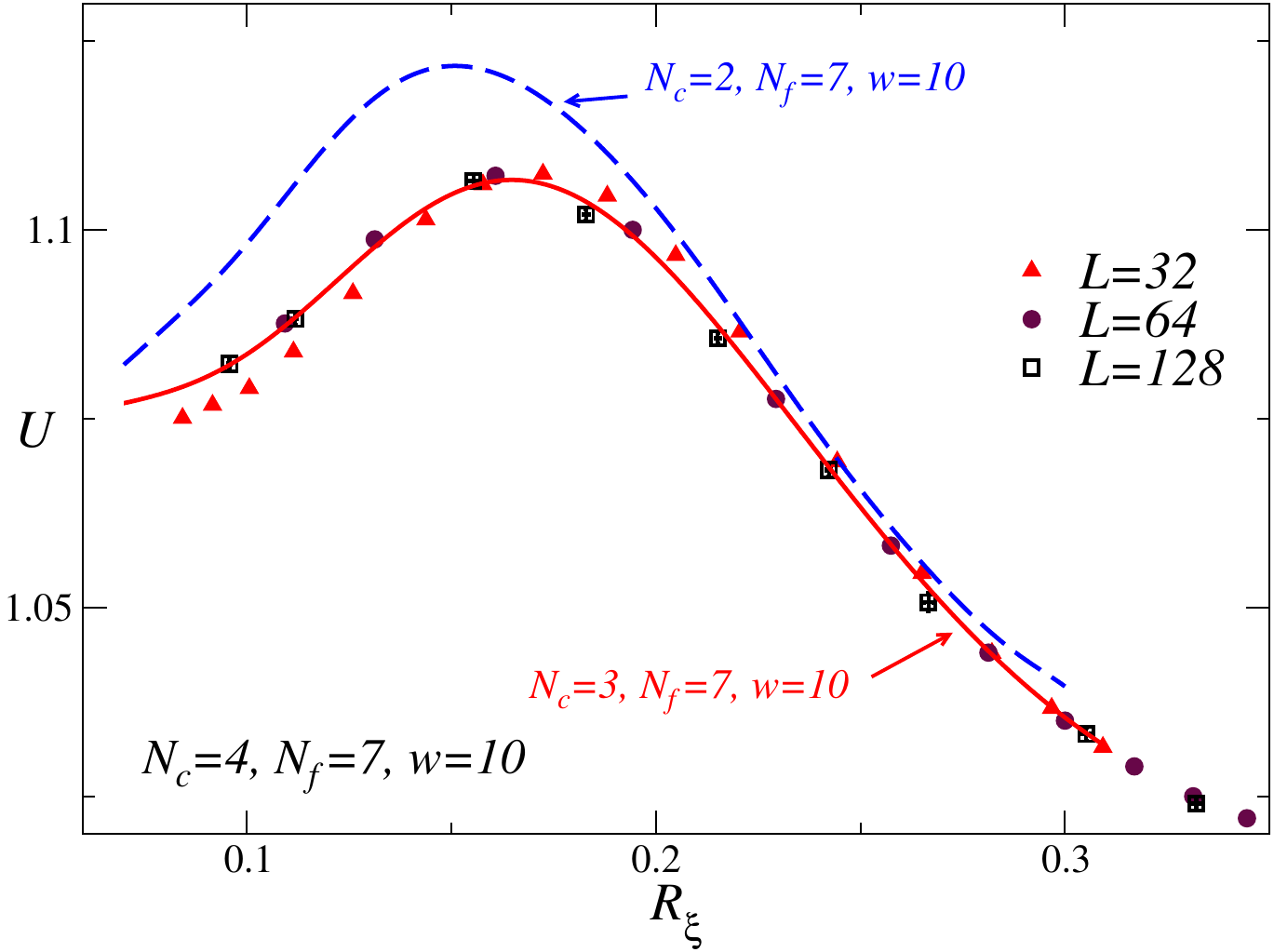}
    \caption{Top: Binder cumulant $U$ versus $R_\xi$ for the case $N_c=3,
N_f=7$ and quartic coupling $w=10$ ($\gamma=0$). The solid line is polynomial interpolation
of the $L=64$ and $L=128$ data. The dashed line is our estimate of
$\mathcal{U}(R_{\xi})$ for the $N_c=2$, $N_f=7$ model.  Bottom: Binder cumulant
$U$ versus $R_\xi$ for the case $N_c=4, N_f=7$ and quartic coupling $w=10$ ($\gamma=0$).
The solid line is our estimates of $\mathcal{U}(R_{\xi})$ for the $N_c=3$,
$N_f=7$ model.}
    \label{fig3c7fw10}
\end{figure}

In Fig.\ref{fig2c7fw10} we present our results for $U(R_{\xi},L)$ in the models
with $N_c=2$ and $N_c=5$, with quartic coupling fixed to $w=10$ and $\gamma=0$.
For $N_c=2$ we observe that FSS corrections are quite small, and results coming
from lattice sizes $L=64$ and $L=128$ are consistent with each other.  We thus
used a polynomial interpolation of $L=64$ and $128$ data to estimate the
universal scaling curve $\mathcal{U}(R_{\xi})$ of this model. This universal
curve is then compared, in the bottom panel of Fig.\ref{fig2c7fw10}, with the
results obtained for the $N_c=5$ model. In this case we observe scaling
corrections larger than the ones obtained for $N_c=2$, however data for the
model with $N_c=5$ are clearly converging to the same asymptotic curve as the
$N_c=2$ model. The different approaches to the asymptotic curve observed for
$N_c=2$ and $N_c=5$ show that color-reflection symmetry is not a generic
symmetry of the model, but an emerging symmetry in the critical domain.

In Fig.\ref{fig3c7fw10} we report data for $U(R_{\xi},L)$ in the models with
$N_c=3$ and $N_c=4$, still with quartic coupling $w=10$ and $\gamma=0$. We can
immediately note that the critical behavior found in this case is different
from the one observed when $N_c=2$ (and $N_c=5$), whose universal curve
$\mathcal{U}(R_{\xi})$ is represented in Fig.\ref{fig3c7fw10} by the blue
dashed curve. This fact provides an indication that color degrees of freedom
actively participate to the critical behavior when $w>0$. We remind the reader
that a completely different behavior was found in Ref.~\cite{BFPV_asymp} for
$w=0$. In that case the low-temperature critical behavior was the same as the
$\mathbb{R}P^{N_f-1}$ model for all the values $N_c>2$ studied. As discussed
in Sec.~\ref{sec_minima} the same is expected to happen for negative
values of the quartic coupling for any $N_c$.

We have thus seen that color-reflection symmetry is realized for $N_c=2$ and 5,
and that this symmetry is not trivially realized in the low temperature phase
(i.e. there is a dependence of the critical behavior on $N_c$). The same
procedure adopted to compare the results obtained for $N_c=2$ and $N_c=5$ can
now be applied to analyze also the cases $N_c=3$ and $N_c=4$. The red solid
curves in Fig.\ref{fig3c7fw10} are obtained by polynomially interpolating data
corresponding to $L=64$ and $128$ with $N_c=3$ (residual corrections to scaling
are visible between $L=64$ and $128$, but they are smaller than two standard
deviations). The same curve is seen to consistently describe the critical
behavior of the model also for $N_c=4$, as expected on the basis of the
emergent color-reflection symmetry.

\begin{figure}[tb]
    \centering
    \includegraphics[width=0.95\columnwidth, clip]{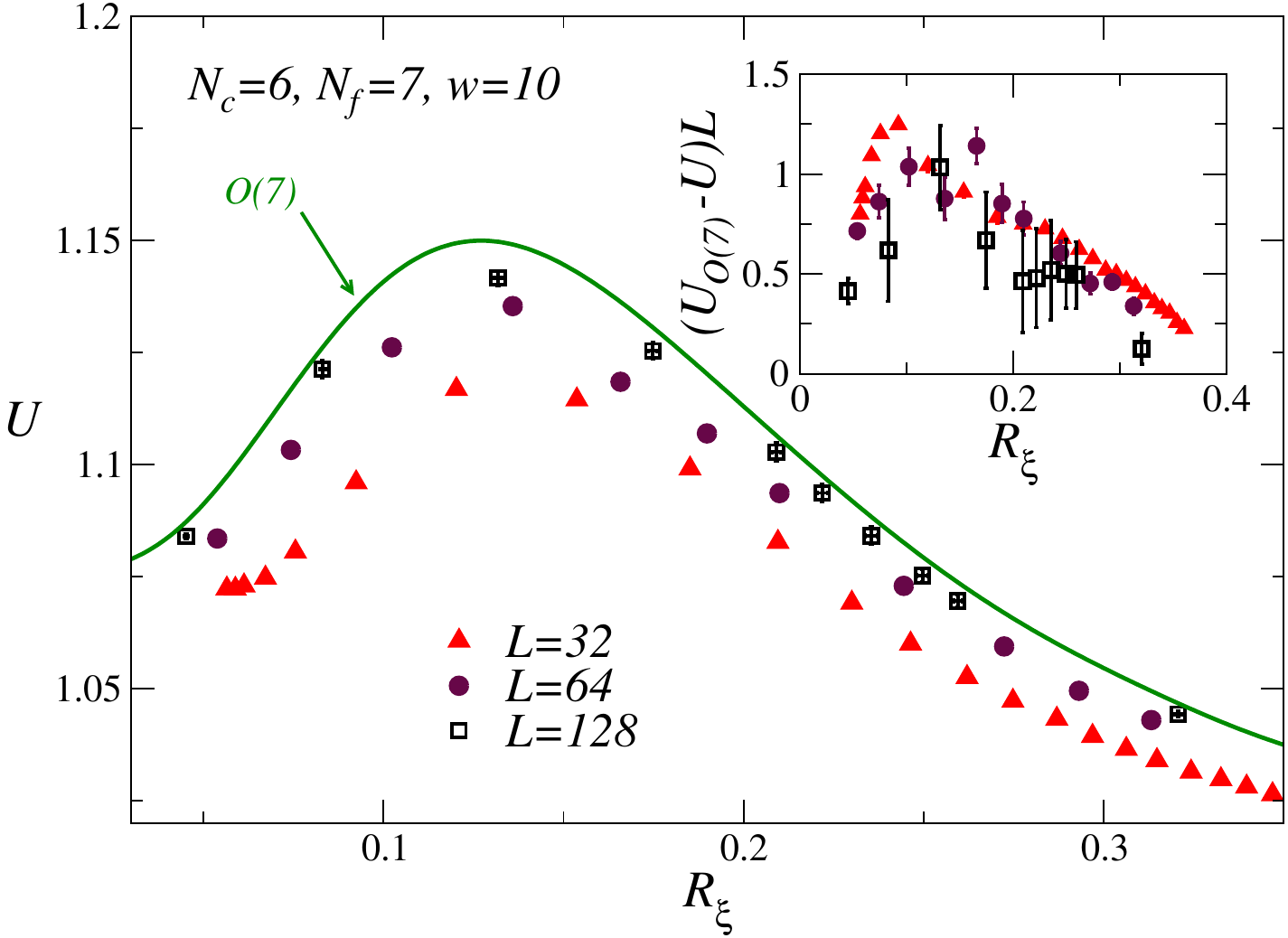}
    \caption{Binder cumulant $U$ versus $R_\xi$ for the case $N_c=6, N_f=7$ and
quartic coupling $w=10$ ($\gamma=0$). For reference we also report our estimate of
$\mathcal{U}(R_{\xi})$ for spin-2 observables in the vector $O(7)$ model,
obtained by a polynomial interpolation of data coming from $L=32$ and $L=64$ lattices
(consistent with each other within statistical uncertainties).  In the inset, scaling
corrections are shown to be roughly consistent with a $L^{-1}$ behavior.}
\label{fig6c7fw10}
\end{figure}

Finally, in Fig.\ref{fig6c7fw10}, we discuss the extreme cases $N_c=1$ and
$N_c=6$, once again for $w=10$ and $\gamma=0$. For $N_c=1$ the model studied
reduces to the standard $O(7)$ model (although we use spin-2 observables
instead of the usual vector ones), and again an emergent color-flavor
reflection symmetry is observed: despite the presence of large scaling
corrections, the critical behavior of $U(R_{\xi}, L)$ for the model with $N_c=6$ is
compatible with the one observed in the $O(7)$ model. As often happens for
$O(N)$ models, corrections to scaling are roughly compatible with a $L^{-1}$
scaling \cite{Balog_thepuz}. The peak value of the Binder
cumulant ($U_{\mathrm{max}}\approx 1.15$) is already sufficient to show that this critical
behavior is different from the one seen for $N_c=2$ ($U_{\mathrm{max}}\approx1.125$) and
$N_c=3$ ($U_{\mathrm{max}}\approx1.10$).

\begin{figure}
    \centering
    \includegraphics[width=0.95\columnwidth, clip]{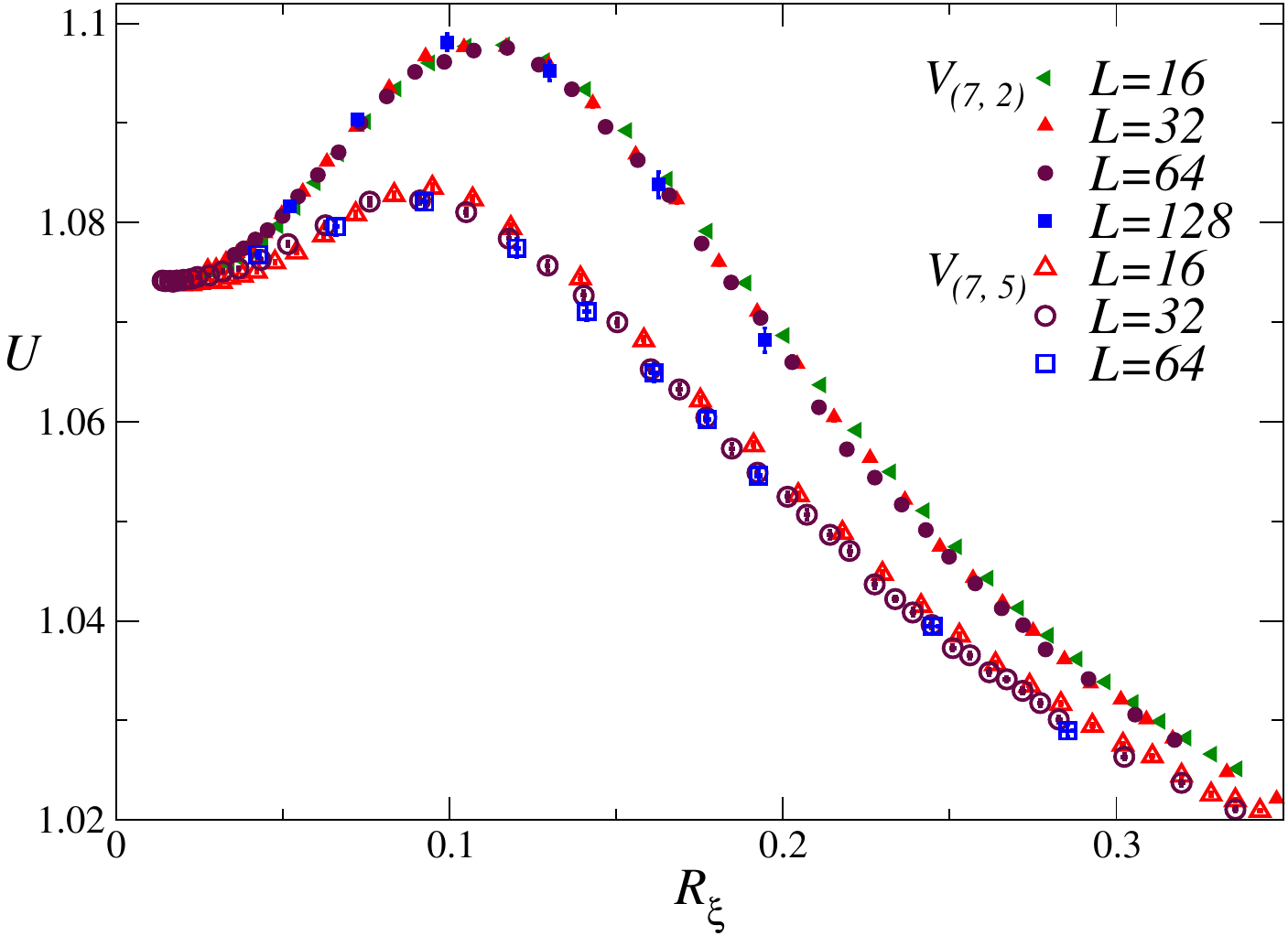}
    \caption{Binder cumulant $U$ versus $R_\xi$ for the Stiefel
models $V_{(7, 2)}$ and $V_{(7, 5)}$.}
    \label{figstiefel2c7f}
\end{figure}

\begin{figure}
    \centering
    \includegraphics[width=0.95\columnwidth, clip]{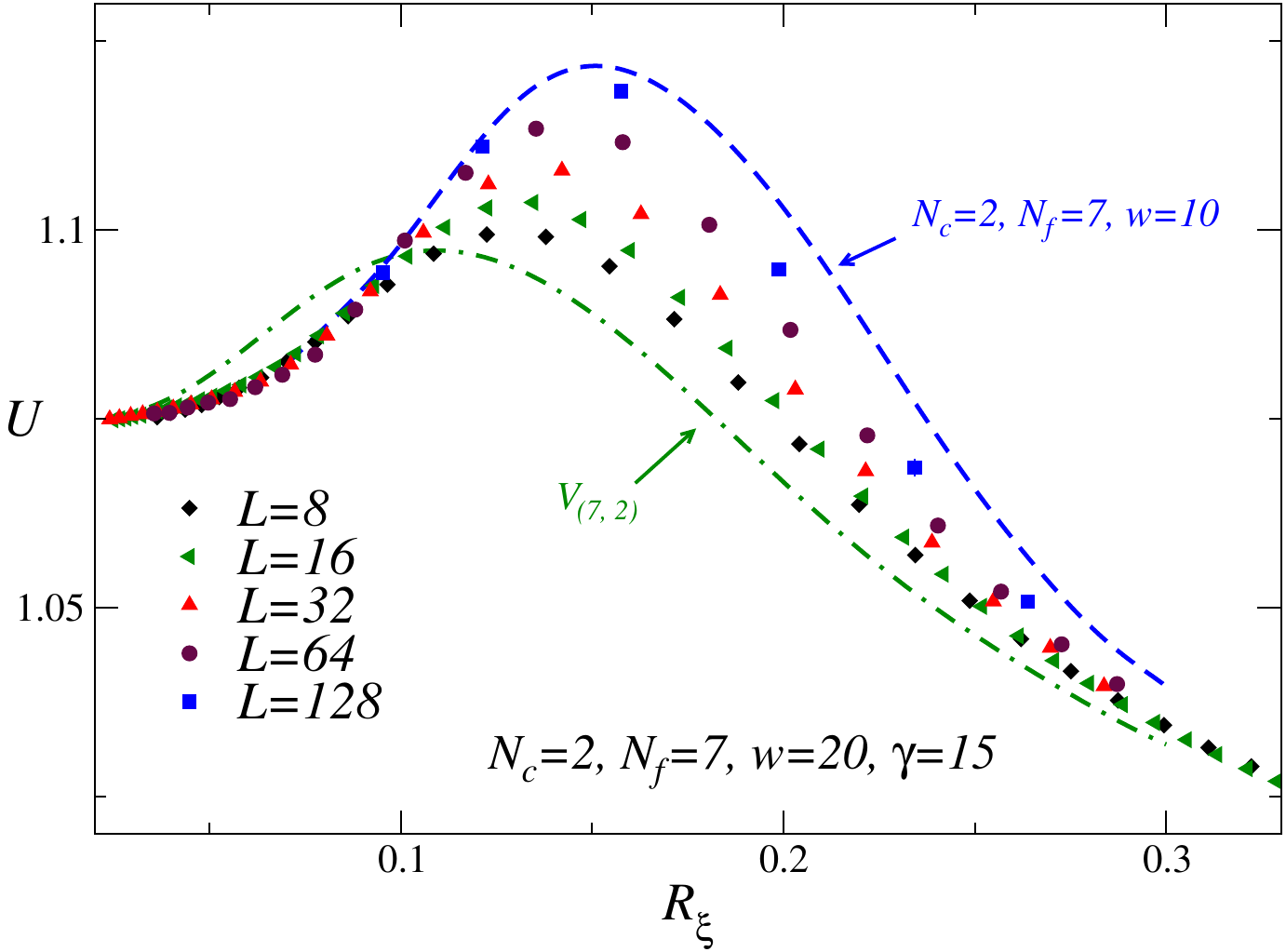}
    \caption{Binder cumulant $U$ versus $R_\xi$ for the case $N_c=2, N_f=7$,
quartic coupling $w=20$ and gauge coupling $\gamma=15$.
The dashed line represents our estimate of $\mathcal{U}(R_{\xi})$ for the model
with $N_c=2$, $N_f=7$, $w=10$ and $\gamma=0$, while the dotted-dashed line is
estimate of $\mathcal{U}(R_{\xi})$ for the Stiefel model $V_{(7,2)}$.}
    \label{fig_crossover}
\end{figure}

We finally want to provide some indication that, for large values of $\gamma$,
data can be significantly affected by the crossover between the Stiefel and the
Grassmanian critical behaviors, with color-flavor reflection symmetry that is
absent in the Stiefel case. For this purpose we have first of all determined
the scaling curves $\mathcal{U}(R_{\xi})$ for the Stiefel models $V_{(7,2)}$
and $V_{(7,5)}$, shown in Fig.~\ref{figstiefel2c7f}. These curves make evident
the fact that the color-reflection symmetry characterizing the critical
behavior of the gauge models is not present in the Stiefel case.  Data for the
gauge model with $N_c=2, N_f=7$, quartic coupling $w=20$ and gauge coupling
$\gamma=15$, are reported in Fig.~\ref{fig_crossover}, and clearly display a
crossover behavior between two different regimes. Results slowly converge to
the expected Grassmanian asymptotic curve when increasing the lattice size, but
on small lattices they are far from it and quite close to the results of the
Stiefel model $V_{(7,2)}$. Note however that on a finite lattice the
$\gamma\to\infty$ limit of the gauge model does not coincide with the Stiefel
model with periodic boundary conditions, due to the presence of nontrivial
holonomies (Polyakov loops) that can not be gauged away. To obtain an exact
matching one should generalize to the non-Abelian case the fluctuating boundary
conditions used in \cite{BFPV_asymp_rpn} for the Abelian case.

\section{Conclusions}\label{sec_conclusions}

In this paper we have addressed the interplay of local and global symmetries in
determining the universal low-temperature critical behavior of 2D scalar
models. In particular, we have considered multiflavor lattice models with
$SO(N_c)$ gauge symmetry and non-Abelian $O(N_f)$ global symmetry, in the
presence of a quartic interaction between the scalar fields, thus extending the
results already presented in \cite{BFPV_asymp}.

By studying the minimum-energy configurations we identified two different
low-temperature regimes. If the coefficient of the quartic coupling is negative
($w<0$), the $SO(N_c)$-gauge models share the same low-temperature critical
behavior of the $\mathbb{R}P^{N_f-1}$ models, and color degrees of freedom do
not play any active role in the critical domain. In particular the
low-temperature effective theory is independent of the number of colors $N_c$.
If instead $w>0$, the nature of the low-temperature regime crucially depends on
the number of colors and flavors. If $N_f\le N_c$ no global symmetry remains,
and no diverging correlation length and critical behavior are present. If
instead $N_f>N_c$, the low-temperature behavior is expected to be described by
the gauged non-linear $\sigma$ model field theory associated with the real
Grassmannian manifold $SO(N_f)/(SO(N_c)\times SO(N_f-N_c)$).

To support the identification with the Grassmanian NL$\sigma$M field theory of
the critical behavior when $N_f>N_c$ and $w>0$, we investigated the
$N_c$-dependence of the asymptotic scaling curve $\mathcal{U}(R_{\xi})$ and, in
particular, the presence of the color-flavor reflection symmetry
$N_c\leftrightarrow N_f-N_c$ in the critical region. For this purpose we
focused on the case $N_f=7$, studied for several values of $N_c$ with $w=10$
and $\gamma=0$. We numerically verified that gauge degrees of freedom do play a
role in the critical behavior, since the asymptotic scaling curves
obtained in the cases $N_c=1,2$ and $3$ are different
from each other. Moreover, we provided robust evidence of the emergence of
color-reflection symmetry in the critical domain, with the results obtained
using $N_c$ and $N_f-N_c$ colors that approach the same asymptotic curve in the
FSS limit.

It is interesting to note that for some values of $N_c$ and $N_f$ the
Grassmanian models can have peculiar properties that will be worth
investigating. For example, the $N_c=2$ model admits instanton solutions
analogous to that of the $\mathbb{C}P^{N-1}$ models, a fact already noted in
the seminal work \cite{Dadda_A1}. However to our knowledge the $\theta$
dependence of this model has never been systematically investigated.

Another aspect that deserves to be further explored is the stability of the results
presented in this paper against an explicit breaking of the gauge symmetry.
It is tempting to guess a gauge breaking term to be relevant for $w>0$ and
irrelevant for $w<0$, based on the analogy with the case of 3D Abelian models
with critical \cite{BPV_lattice} or non-critical \cite{BPV_breaking} gauge
modes.  It is however not clear a priori how far this analogy can be trusted,
also because in a non-Abelian theory we have much more freedom on the form of
the gauge symmetry breaking term, that could for example leave a residual
continuous subgroup exact. 

Finally, in this work we focused on real scalar models with orthogonal gauge
symmetry, but an extension to the case of complex scalar models with unitary
symmetry can be immediately carried out. In this latter case we can have,
depending on the sign of the quartic interaction, a $\mathbb{C}P^{N_f-1}$
critical behavior (for $w<0$) or a complex Grassmannian $SU(N_f)/(SU(N_c)\times
SU(N_f-N_c))$ critical behavior (for $w>0$). 

\noindent
\textit{Acknowledgement}. Numerical simulations have been carried out on the CSN4
cluster of the Scientific Computing Center at INFN-PISA

\appendix 
\section{Monte Carlo simulations} \label{appendix}

In this appendix we present some technical details regarding our Monte Carlo
simulations, the algorithms adopted, and the statistics accumulated. 

For the gauge models the typical statistics used is of the order of $10^7$ 
``complete lattice sweeps'' (defined in the following subsection), 
while for the Stiefel models it is of the order of $10^6$ ``complete lattice
sweeps''. To analyze data and estimate error bars we used standard blocking and
jackknife techniques, and the maximum blocking size adopted was of the order of
$10^5$ and $10^4$ data, for gauge and Stiefel models respectively.

For lattice models with local gauge symmetry, we observed that the value of the
inverse temperature $\beta$ has to be increased to keep $R_{\xi}$ constant
while increasing the value of $N_c$ (at fixed $N_f=7$). Specifically, for the
lattice size $L=128$, we used for the model with $N_c=2$ inverse temperatures
in the range $\beta \in [1.7,2.1]$, for $N_c=3$ we used $\beta\in [3.0,3.5]$, for $N_c=4$ we used $\beta
\in [4.5, 5.0]$, while for the case $N_c=6$ the range  $\beta \in [7.5, 8.3]$
was adopted. For the case $N_c=5$ our largest lattice was $L=64$, and in this case
we used $\beta\in [5.8, 6.3]$. Using these inverse temperature intervals we got
values of $R_{\xi}$ in the range $[0.05,0.3]$ in all the cases.

\subsection{Algorithms: gauge models}

To update the field $\Phi_{\bm x}$ of the gauge models we use both a Metropolis
\cite{MRRTT_equat} and a pseudo-overrelaxation algorithm \cite{A_anover, C_overr}. In the
Metropolis update, the trial field $\Phi^\prime_{\bm x}$ is generated from
$\Phi_{\bm x}$ by rotating two random matrix elements by an angle drawn from an
uniform distribution in $[-\alpha,\alpha]$. The value of $\alpha$ is chosen in
such a way to have an average acceptance rate of about $30\%$.  The
pseudo-overrelaxation step is performed by using as trial field the reflection
of $\Phi_{\bm x}$ with respect to the force $F_{\bm x}$ defined by
\begin{equation}
    F_{\bm x} \equiv \sum_{\mu} \bigg( U_{\bm x, \mu} \Phi_{\bm x + \mu} 
    + U^t_{\bm x - \mu, \mu}\Phi_{\bm x-\mu}\bigg)\,,
\end{equation}
thus
\begin{equation}
    \Phi_{\bm x}^\prime = \frac{2 \Tr \big(\Phi^t_{\bm x}F_{\bm x}\big)}{\Tr 
\big(F^t_{\bm x}F_{\bm x}\big)}F_{\bm x} - \Phi_{\bm x}\,.
\end{equation}
This trial field is then accepted or rejected by a Metropolis test, which is
unnecessary when $w=0$ since in that case the update is energy preserving.

The gauge field $U_{\bm x, \mu}$ is updated using the Metropolis algorithm,
with the trial link $U^\prime_{\bm x, \mu}$ generated by applying to $U_{\bm x,
\mu}$ a random $SO(2)$ matrix randomly embedded in $SO(N_c)$. Also in this case the
parameters of the rotation are tuned in order to maintain an average acceptance
of approximately $30\%$.  

A “complete lattice sweep" is defined to be a series of 10 update sweeps 
on the whole lattice for both scalar and gauge fields. For the gauge field the
Metropolis update is always adopted, while for the scalar field a single Metropolis
update is followed by 9 pseudo-overrelaxation steps.

\subsection{Algorithms: Stiefel models}

The Stiefel Hamiltonian is defined in Eq.(\ref{ham_stiefel}) by using the field
$\pi_{\bm x}$, which is a $N_c\times N_f$ matrix subject to the constraint
$\pi_{\bm x}\pi_{\bm x}^t=\mathbb{1}_{N_c\times N_c}$.  Assuming for the sake of the
simplicity that $N_f>N_c$, we represent $\pi_{\bm x}$ by the first $N_c$ rows
of the $N_f\times N_f$ orthogonal matrix $\widetilde{\pi}_{\bm x}$, that will
be our fundamental field in the following.  The update of $\widetilde{\pi}_{\bm
x}$ is performed by using the Metropolis algorithm \cite{MRRTT_equat} and the
single cluster algorithm \cite{W_collec}.

The Metropolis update of the field $\pi_{\bm x}$ is performed by using the trial
state $\widetilde{\pi}^\prime_{\bm x}=\widetilde{\pi}_{\bm x} O$, where $O$ is
an $O(2)$ random rotation embedded in a random way in $O(N_f)$. The rotation
angle is drawn from a uniform distribution in $[-\delta, \delta]$, and the
value of $\delta$ is chosen to obtain an average acceptance of about $30\%$. 

In the cluster update we start by generating the random unit-length
$N_f$-vector $v^{\alpha}$ and selecting a random lattice site. The cluster construction is 
performed by activating the link $\bm x - \bm y$  
with probability $p_{\bm x, \bm y}$ given by
\begin{equation}
  p_{\bm x, \bm y}=1-\exp\bigg(\min(0, 2\beta N_f X)\bigg)\ ,
\end{equation}
where
\begin{equation}
  X=\sum_{i=1}^{N_c}\bigg[\bigg(\sum_{\alpha=1}^{N_f}\pi_{\bm x}^{i\alpha}v^{\alpha}\bigg)
  \bigg(\sum_{\beta=1}^{N_f}\pi_{\bm y}^{i\beta}v^{\beta}\bigg)\bigg]\,.
\end{equation}
The whole cluster is then ``flipped'' using 
\begin{equation}
 (\widetilde{\pi}_{\bm x})^{\alpha\beta} \to (\widetilde{\pi}_{\bm x})^{\alpha\beta} - 
 2v^{\beta}\sum_{\lambda=1}^{N_f}(\widetilde{\pi}^{\alpha\lambda} v^{\lambda})\,,
\end{equation}
which is easily seen to be an $O(N_f)$ matrix.

A ``complete lattice sweep'' is a series of 10 Metropolis updates on the whole
lattice, each one followed by a cluster update.

\end{document}